# Photothermomechanically Efficient, Low-Cost, High-Cycle-Life, Hybrid MXene-Polymer Actuators

Ken Iiyoshi,[1,6]\* Georgios Korres,[1] Orsolya Nagy,[2] Gabriel Roldán,[1] Panče Naumov,[3,4] Stefan Schramm,[5] and Mohamad Eid[1,6]\*

[1]*Division of Engineering, New York University Abu Dhabi, Saadiyat Island, Abu Dhabi, PO Box 129188, United Arab Emirates*
[2]*School of Engineering, EPFL, 1015 Lausanne, Vaud, Switzerland*
[3]*Smart Materials Lab, New York University Abu Dhabi, Saadiyat Island, Abu Dhabi, PO Box 129188, United Arab Emirates*
[4]*Molecular Design Institute, Department of Chemistry, New York University, 100 Washington Square East, New York, NY 10003, USA*
[5]*Applied Organic Chemistry, University of Applied Sciences Dresden/HTW Dresden, Dresden, 01069, Germany*
[6]*NYU Tandon School of Engineering, New York University, Brooklyn, New York, NY 11201, USA*
\*Corresponding authors. Email for correspondence: ken.iiyoshi@nyu.edu (K. I.); mohamad.eid@nyu.edu (M. E.)

## Abstract

Photothermomechanical polymer film actuators stand out among the dynamic components available for soft robotics due to a combination of assets, such as capability for rapid energy transduction, wireless control, and ease of miniaturization. Despite their anticipated superior performance, several design challenges remain. These include high operational temperatures, inadequate mechanical output relative to the radiation energy provided, limited durability during repeated use, and high production costs; such factors hinder the scalability of these actuating materials in practical applications. Here, we report a viable solution by substituting performance-enhancing nanoparticles with MXenes—carbon-based, two-dimensional materials known for their theoretical photothermal conversion efficiency of up to 100%. This led to the development of MXene-dispersed polymer trilayer actuators (MPTAs). Extensive photothermal and thermomechanical characterization demonstrated superior performance compared to previously reported actuators, with a reduced shed power demand (0.1 mW cm$^{-2}$ °C$^{-1}$), substantial bending capacity per irradiation power per time (0.1° mW$^{-1}$ cm$^2$ s$^{-1}$), and enhanced cyclic longevity, with fatigueless operation of at least 1,000 cycles. We demonstrate three applications: A kirigami-inspired flower, parallel manipulator, and soft gripper. Additionally, these materials are cost-effective; thus, they are the optimal choice for long-term, reversible operation with efficient heat-to-work transduction.

## Introduction

One of the core challenges of contemporary robotics is the development of new materials and fabrication schemes that are optimized for delivery across various performance indices (*1*). Over the past decade, soft robotics has established itself as a blooming robotics subfield (*2*), with soft robots set to realize enhanced practicality and broader impact within both consumer and industrial contexts; additionally, soft robotics are the subject of active exploratory efforts on several trajectories (*3*). One such direction focuses on innovative soft actuators for untethered soft actuation, preferably constructed from soft and compliant materials (*4*). This approach addresses the inherent automation challenges associated with conventional soft robots (*5*), and could augment applications that encompass object manipulation (*6*), locomotion (*7*), and rehabilitation (*8*).

Stimuli-responsive polymers represent a category of untethered soft actuating materials, ranging from hard plastics to soft liquid crystal elastomers (LCEs) and hydrogels. These materials respond to untethered external stimuli, including thermal energy, electromagnetic radiation across a wide range of spectra, variations in pH or humidity, magnetic fields, and electric fields (*9*). Among them, thermomechanical polymers driven by electromagnetic radiation are of particular interest (*10, 11*), especially due their applicability in specific situations, such as orchestrated control of actuator swarming, or ability to monitor via thermal transduction (*12*). While polymeric materials generally exhibit expansion upon thermal excitation, particularly within the elastomer subclass, metamaterial design can additionally engender hygroscopic contraction, specifically manifested as dimensional reduction upon evaporative dissipation of adsorbed water vapor (*13, 14*). When polymers are configured into thin-film geometries, they exhibit rapid thermal modulation and actuation kinetics (*15*); thus, polymer-based light-responsive film materials have been subjected to extensive performance assessments (*16, 17*). Despite these developments, however, comparatively modest photothermal and photomechanical conversion efficiencies, significant material costs, and limited longevity in cyclic operational mode remain the primary impediments associated with their implementation. Alternative approaches include the application of coatings or mixtures or the inclusion of one-dimensional (1D) or two-dimensional (2D) photothermal nanoparticles (*18, 19*); nevertheless, further innovations in photothermomechanical polymer materials, potentially through novel compositional strategies, are required to address these challenges, impeding practical deployment.

A family of titanium carbides, formulated as $Ti_3C_2$ (or $Ti_3C_2T_x$) and more commonly referred to as MXenes, has recently garnered considerable attention for their unprecedented photothermal conversion properties (*20, 21*). These materials present a diverse spectrum of applications, ranging from soft robotics (*22*) and energy harvesting (*23*) to optical waveguide arrays (*24*). Based on localized surface plasmon resonance, MXenes has theoretical photothermal conversion efficiencies that approach unity (100%) (*25*), and have low emissivity within the mid-infrared (mid-IR) spectral range (*26*). Expectedly, MXenes surpasses the thermal performance of blackbody materials (*26*).

For actuators, MXenes have been used as mixtures (*27, 28*) and coatings (*29, 30*). The prevalent approach is to prepare a bimorph of pure MXene film and thermomechanical polymer. Upon lighting, the bimorph bends from asymmetric photothermal expansion. The light sources are typically near-infrared (NIR), and solar irradiation or wider infrared (IR) range to a lesser extent. Notably, Cai *et al*. demonstrated the integration of the photothermal properties of MXenes with the thermohygroscopic behavior of cellulose-polycarbonate layers to induce contraction and bending upon NIR irradiation (*31*). The authors formulated a model for the curvature of the film actuator as a function of its layers' thermal expansion, hygroscopic contraction coefficient, and film temperature (*31*). In another notable recent study, Cho *et al*. reported the bending of a MXene-aligned liquid crystal elastomer (LCE) by NIR irradiation (*29*). However, using MXenes as pure films is known to decrease their photothermal conversion efficiency to levels as low as 20% (*30*). Additionally, none of these studies utilized bimorph materials that concurrently exhibit both high expansion and high contraction characteristics, as has been demonstrated with non-MXene actuators (*32, 33*), presumably to enhance bi-directional actuation (*34*); in fact, research on MXene-based film actuators has generally been limited to bilayer structures of a MXene layer and either an expanding or a contracting support layer (*35*). Finally, no prior work has explored the performance of *UVA-driven* MXene actuators, despite its advantages over vis-NIR-IR spectrum. In particular, MXenes have highest absorption in of typically 2.0 absorbance unit (a.u.) over 1.0 in the NIR range (*36-39*), UVA's shorter wavelength enables higher spatial resolution for controlling arrays of actuators with complex lighting patterns (*40-42*), and introducing UVA-based applications to environments that use visible lighting, NIR-based remote controllers, and IR heaters, will not interfere with each other (*40, 43, 44*). Hence, the MXene actuator literature has three notable gaps that require attention: a lack of effective physical MXene dispersion, a lack of combined expansion-contraction bimorph architectures, and a lack of investigation into UVA inputs.

A preliminary effort to mitigate these challenges has been undertaken in our prior research (*45*). Initially, upon being subjected to a limited number of UV irradiation cycles, free-standing film actuators that were 10 mm in length and featured an expanding and contracting bimorph architecture exhibited vertical displacements of up to 0.6 mm at a temperature close to 45 °C. This performance was consistently observed across a range of film widths between 1 and 7 mm. With the incorporation of an intermediate layer, the resultant film manifested a decreased incidence of fabrication imperfections; however, it maintained a comparable actuation displacement of 0.6 mm. Following the incorporation of 2.5% (vol) MXene within a constituent layer of the trilayer construct, an increase of approximately 3.5 °C in the actuation temperature and an increase of approximately 30% in the bending were observed. Furthermore, a larger analog, 29 mm in length, demonstrated a maximum force exertion of 10 mN and an average force output of 7.9 mN. However, the degree to which the contributions from UV irradiation, MXene particulate dispersiveness, and the configuration of the trilayer are leveraged remain indeterminate in the absence of comprehensive photothermal and thermomechanical characterization.

Toward these objectives, the present study proposes UV-induced actuation of MXene-dispersed polymer trilayer actuators (MPTAs), wherein MXene powder is dispersed in an elastomer, which is layered onto substrates such as plastics and paper. Here, the conceptual architecture of the MPTA is presented and substantiated by photothermal characterization of the efficacy of the MXene-dispersed elastomer, a well-defined trilayer interface, and an optimized MXene concentration. Thermomechanical characterization revealed that an MPTA with 2.5% MXene within the elastomeric layer has decreased irradiation requirements, reduced material expenditure, substantial bending capacity, and enhanced cyclic durability in comparison to the antecedent architectures. To exemplify its potential, the MPTAs were assembled into a kirigami-inspired flower-shaped (KIFS) design, which demonstrated consistent bending and actuation over a period exceeding 60,000 s, equivalent to 1,000 actuation cycles. This superior performance was further corroborated by thermographic analysis, displacement measurements, modulus and stiffness measurements, and microscopic imaging, which confirmed the sustained structural integrity of the MPTA upon prolonged cyclic operation. We show a parallel manipulator and soft gripper demonstration as promising applications. Collectively, these findings delineate a novel trajectory for the advancement of MPTAs via the combination of a novel approach for MXene integration in structural and host materials that potentially surpasses the performance of other photothermomechanical film actuators for applications in soft robotics.

## Results

*Conceptual architecture*

The conceptual architecture of the MPTA is delineated in Figure 1a. Under ambient conditions (room temperature), the film assumes a planar configuration. Upon irradiation with UV light, the film is uniformly heated, whereupon the top layer and (to a lesser extent) the intermediate layer expands, while the bottom layer contracts. After the heat dissipates naturally, the MPTA flattens to its original shape. A scanning electron microscopy (SEM) image of a cross-section of the sagittal plane of the MPTA showed a trilayer structure, with thicknesses of approximately 70 μm, 50 μm, and 100 μm for the elastomeric, plastic, and paper layers, respectively (Figure 1b). Figure 1c shows that that the MXenes are well dispersed. The (EDS) chemical elemental map in Figure 1d confirms the presence of titanium-based nanosheets, and Figures 1e and 1f show the presence of carbon and oxygen throughout the film, consistent with the chemical composition of MXenes. The EDS spectrum and atomic percentages are in Supplementary Table 1.

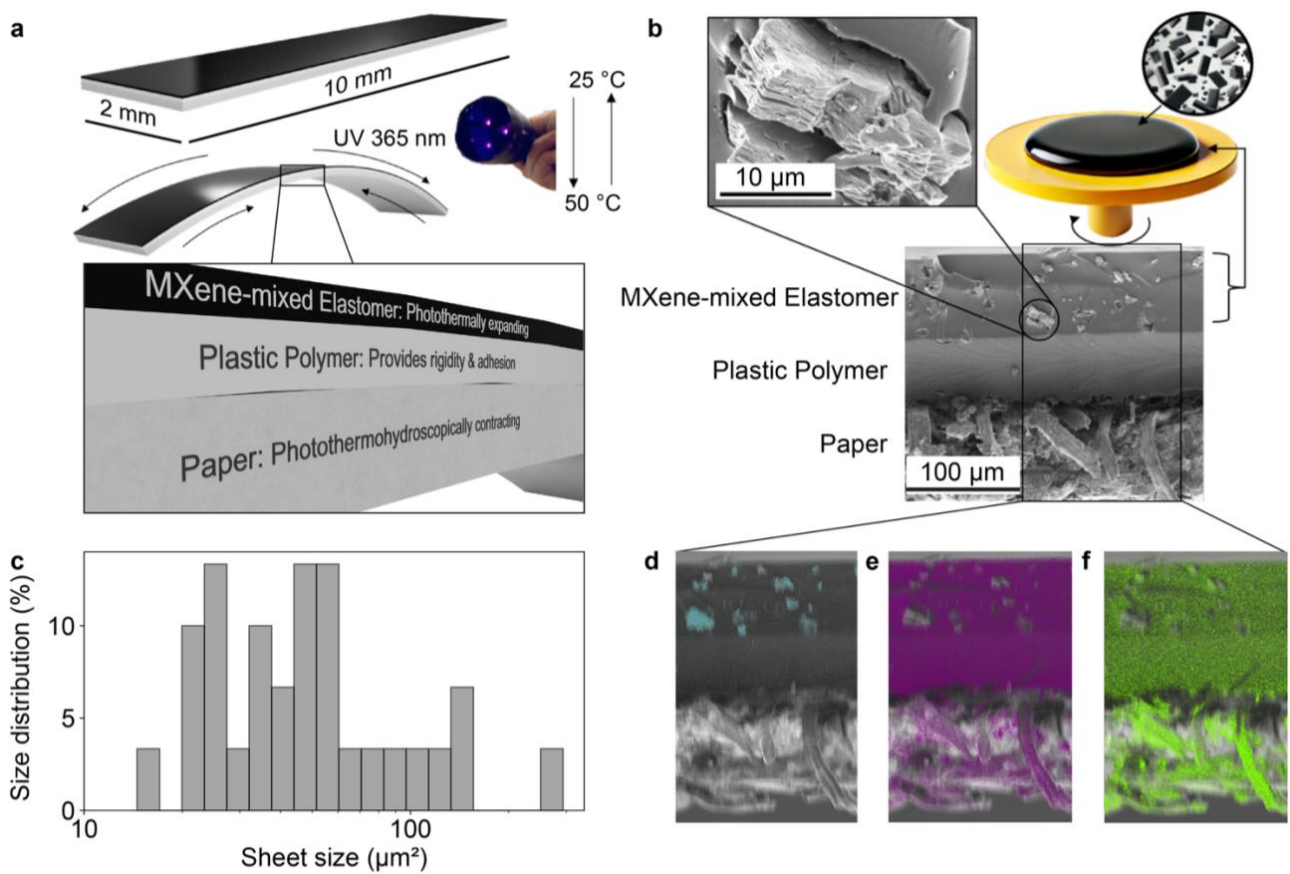

**Fig. 1 | Preparation and structure of MPTA. a** Schematic of the bending mechanism. The actuator maintains a planar configuration in its quiescent state at ambient temperature. Upon exposure to UV light, the surface layer of the MXene-mixed elastomer undergoes isotropic expansion, while the paper substrate, as a bottom layer, undergoes anisotropic contraction, predominantly along its longitudinal axis, resulting in macroscopic bending. The intermediate plastic layer ensures interfacial adhesion between the expansive

and contractive layers. **b** Representative cross-sectional scanning electronic microscopy (SEM) image of MPTA, revealing the trilayer composition: a top layer consisting of a UV-cured, spin-coated MXene-dispersed photopolymer, an intermediate layer of spin-coated plastic photopolymer, and a bottom layer of paper. The alignment of the paper fibers is approximately orthogonal to the depicted cross section. A typical MXene nanosheet with its characteristic accordion-like structure is highlighted in the inset. **c** Statistical distribution of the size of the MXene sheets with respect to the cross-sectional area. **d-f** Pseudocolored chemical elemental maps corresponding to the region of the SEM image depicted in panel b with titanium shown in cyan **d**, carbon in purple **e**, and oxygen in light green **f**.

The top layer constitutes a photothermally expansive MXene-dispersed elastomer. A zoomed-in image of this layer reveals its characteristic accordion-like multilayered microstructure typical of MXenes. This dispersion was prepared as an admixture of particulate MXene with an elastomer precursor resin, followed by deposition via spin-coating and subsequent *in situ* photopolymerization. The concentration of MXene relative to the elastomer was 2.5%. After sieving, the size of the nanosheets in the particulates ranged from approximately 10 to 300 $\mu m^2$ (Fig. 1c), and the particles were homogeneously dispersed without discernible agglomeration. An elastomer was used for the expansive layer due to its low stiffness (*46*), which ensures flexibility. This property renders elastomers more susceptible to thermally induced separation and consequently accounts for a high coefficient of thermal expansion (CTE) within the range of $100–1,000·10^{-6}$ $K^{-1}$ (*46*).

The intermediate layer is composed of plastics and augments the interfacial adhesion between the top and bottom layers. In contrast to the more viscous elastomer resin, the plastic resin exhibits enhanced penetration into the paper substrate, and this property decreases the propensity for delamination upon photopolymerization. Analogously, MXenes are incorporated within the elastomeric layer rather than within the plastic or paper layers to minimize delamination (*45*). The inclusion of MXenes has been shown to attenuate the interlayer adhesion strength, as the intermolecular hydrogen bonds formed by MXene surface functionalities are energetically weaker than the intramolecular covalent bonds established within polymer layers during photopolymerization (*31*).

The paper layer plays the role of a thermally contracting component. In contrast to the prevalent behavior of most materials that thermally expand (*46*), conventional copy-quality paper undergoes hygroscopic contraction along the axis orthogonal to the alignment of cellulose fibers imparted during its fabrication process (*47*). Due to its high cellulose content, paper has a relatively large hygroscopic expansion coefficient (CHE) of up to 0.1 $C^{-1}$, where C denotes the moisture concentration of paper; this phenomenon enables pronounced dimensional contraction in response to variations in ambient humidity (*48*). As the temperature increases, the amount of moisture adsorbed by the paper's hydrophilic functional groups decreases linearly, inducing contraction of the cellulose fibers (*31*). Additionally, the paper exhibits a notably low CTE, typically in the range of $4–16·10^{-6}$ $K^{-1}$, ensuring exceptional dimensional stability under thermal fluctuations (*48*). Consequently, when integrated with elastomeric polymers, which typically have CTEs up to an order of magnitude greater than that of paper (*49*), the resulting trilayer structure has, expectedly, a higher bending capacity than that of its antecedents.

### *Photothermal characterization*

Since photothermal characterization is one of the key factors associated with enhanced performance, we first demonstrated the effect of dispersion of the MXenes in the MPTA. The presence of MXene particulates in the MPTA was confirmed through Fourier transform infrared (FT-IR) spectroscopy (Figure 2a). By quantifying the transmittance of MPTAs with varying MXene concentrations across a range of wavenumbers, we observed pronounced attenuation valleys in the transmittance. These valleys correspond to the spectral features characteristic of the chemical fingerprint of MXenes within the 1400–400 $cm^{-1}$ region, as referenced against published FT-IR spectra of MXenes (*50*). The characteristic spectral features include stretching of carbon–fluorine (C–F) bonds at 1100 $cm^{-1}$ (*51*), titanium–fluorine (Ti–F) bonds in the 750–700 $cm^{-1}$ range (*50*), titanium–oxide (Ti–O) bonds in the 650–550 $cm^{-1}$ range (*51*), and titanium carbide (Ti–C) bonds in the approximately 450–350 $cm^{-1}$ range (*50*). Peaks outside these regions further corroborate the presence of MXenes, although they are not solely attributable to the constituent polymers and could also be derived from the terminal functional groups of MXenes. One such peak at approximately 1640 $cm^{-1}$ is associated with water adsorbed on MXenes (*29, 31*). Another indicator was the presence of hydroxyl OH functional groups, which was evident through spectral bands observed at approximately 3335 $cm^{-1}$ (*31*) and 1395 $cm^{-1}$ (*29*). Additional spectral peaks are attributed exclusively to polymeric constituents, such as the carbon double bond stretching C=C in the 3100–3010 $cm^{-1}$ region and the carbonyl (C=O) bond stretching at approximately 1750–1700 $cm^{-1}$ (*52*).

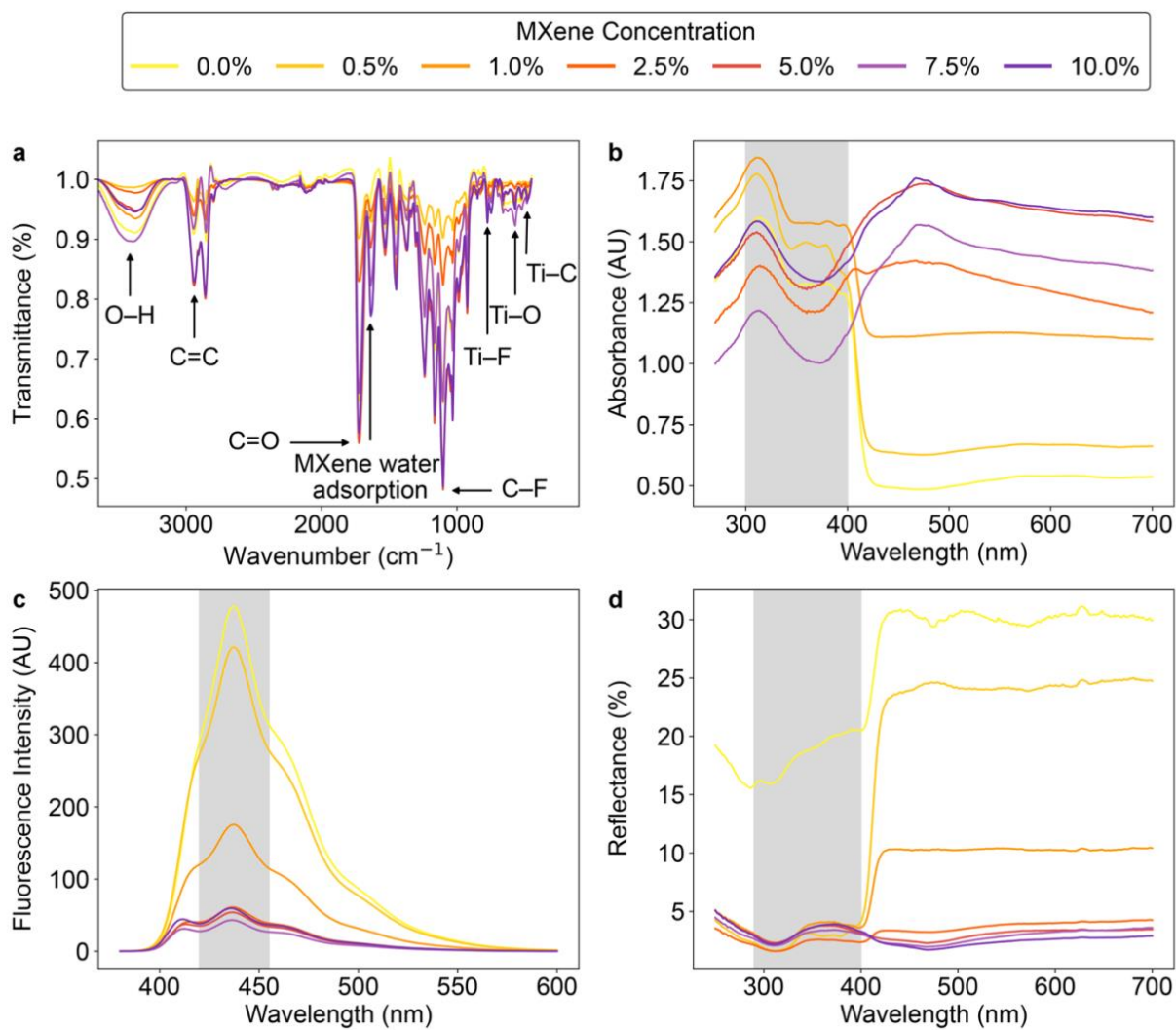

**Fig. 2 | MPTA photothermal characterization. a** FT-IR spectra revealing an attenuation valley at approximately 1640 cm$^{-1}$ that increases with increasing MXene concentration. Other salient spectral features, indicative of MXenes and polymers, are also highlighted, including stretching bands from hydroxyl (O–H), carbon–carbon double bond (C=C), carbonyl (C=O), carbon–fluorine (C–F), titanium–fluorine (Ti–F), titanium–oxygen (Ti–O), and titanium–carbide (Ti–C) bonds. **b** Absorbance spectra in the ultraviolet–visible (UV–vis) spectral range. The optimal photothermal effect is observed at a MXene concentration of 2.5% within the elastomeric layer, demonstrating pronounced absorption in both the UV region, within the spectral region of interest (ROI), and within the visible light spectrum. **c** Fluorescence spectra in the UV–vis spectral range. MPTA incorporating 2.5% MXene exhibited minimal fluorescence emission within the spectral region of interest (ROI), corresponding to the blue light range. **d** Reflectance spectra in the UV–vis spectral range. The MPTA incorporating 2.5% MXene has negligible reflectance in the UV region, within the spectral region of interest, and within the visible region.

For most MXene-associated peaks, a decrease in transmittance is observed with increasing MXene concentration. Figure 2a illustrates that the attenuation valley is enhanced above 2.5% MXene, with 2.5% of the MXene in the elastomeric layer being an optimal composition. Upon irradiation of MPTA with UV light, the incident energy is partitioned into photothermal conversion, fluorescence emission, and reflected radiation. A higher photothermal efficiency is contingent upon maximizing the absorbance while minimizing both the fluorescence and the reflectance. Figure 2b, which shows the absorbance characteristics of MPTAs as a function of the MXene concentration, reveals a positive correlation between the MPTA light absorbance profile and the MXene concentration. The relationship between the UV–vis-IR spectrum-wide absorbance and the corresponding absorbed irradiation power is given by Eq. 1 (*36*):

$$P = P_{in}(1-10^{-A}) \quad (1)$$

where $P$, $A$, and $P_{in}$ represent the absorbed radiant power, MXene-polymer light absorbance, and input irradiation power, respectively. The absorbance, $A$, is further defined by Beer–Lambert's law (Eq. 2):

$$A = -log_{10}(P_{out}/P_{in}) \quad (2)$$

where $P_{out}$ denotes the transmitted radiant power. The UV–vis absorption remains consistently enhanced across all the MXene concentrations investigated. This is attributed to the elastomeric and plastic photopolymer, as well as the paper's UV-absorbing property (*53*), combined with the MXene's dark color. More importantly, beginning at 2.5%, a synergistic effect was observed, characterized not only by pronounced UV absorption but also by substantial visible light and absorption from 400 to 700 nm. This result suggests that any fluorescent emission originating from UV fluorescent pigments mixed into the intermediate plastic layer of MPTA may be effectively captured and converted into thermal energy.

Figure 2c shows that the incorporation of at least 2.5% MXene results in a discernible reduction in energy dissipation from the MPTA via fluorescence emission. This suggests that MPTA absorbs UV light in the form of thermal energy instead of dissipating through fluorescence. Indeed, a MXene concentration of 2.5% correlates with the visual transition of the MPTA toward a darker hue, indicative of enhanced thermal energy absorption. Additionally, the reflection spectra depicted in Figure 2d demonstrate that for MXene concentrations equal to or exceeding 2.5%, a reduced proportion of incident visible light is reflected by the MPTA. Instead, it is posited that the incident radiation is predominantly absorbed by the MPTA, consistent with its darker chromaticity. Thus, the incorporation of a minimum of 2.5% MXene facilitates the absorption of visible light that would otherwise be either reflected by the MPTA or absorbed by the fluorescing photopolymer, with MXenes serving as the primary receptors for optical absorption.

Collectively, in conjunction with Eq. 1, these results indicate that MPTAs with MXene concentrations exceeding 2.5% exhibit comparable photothermal conversion efficiencies and demonstrate superior photothermal conversion efficiency relative to those with lower MXene concentrations (i.e., 0%, 0.5%, or 1%). These observations, together with the subsequent temperature data, provide compelling justification for the selection of an actuator with 2.5% MPTA as the optimal temperature for practical application.

***Thermomechanical characterization***

MPTAs incorporating varying MXene concentrations were exposed to UV radiation, 500 s light and dark conditions to evaluate the enhancement of photothermal conversion capabilities (Supplementary Fig. 2a). While only marginal temperature changes—on the order of a few °C—were discerned between the 2.5% and 5% MPTAs, a general trend of temperature increase with increasing MXene concentration at the peak steady state was consistently observed. Figure 3a comprehensively summarizes these findings, illustrating an 18 °C disparity in peak temperature between the MPTA devoid of MXenes and the MPTA with the highest MXene concentration (10%), which attained maximum temperatures of 62 °C and 80 °C, respectively. For concentrations between 0% and 5%, each MPTA showed a temperature increase of approximately 1 °C compared to the previous concentration, except for the 2.5% MPTA, which reached 66.9 °C, and 5% MPTA, which reached 65.2 °C. A further increase in the MXene concentration to 7.5% resulted in an additional 8.9 °C increase to the temperature, resulting in a peak temperature of 74.1 °C.

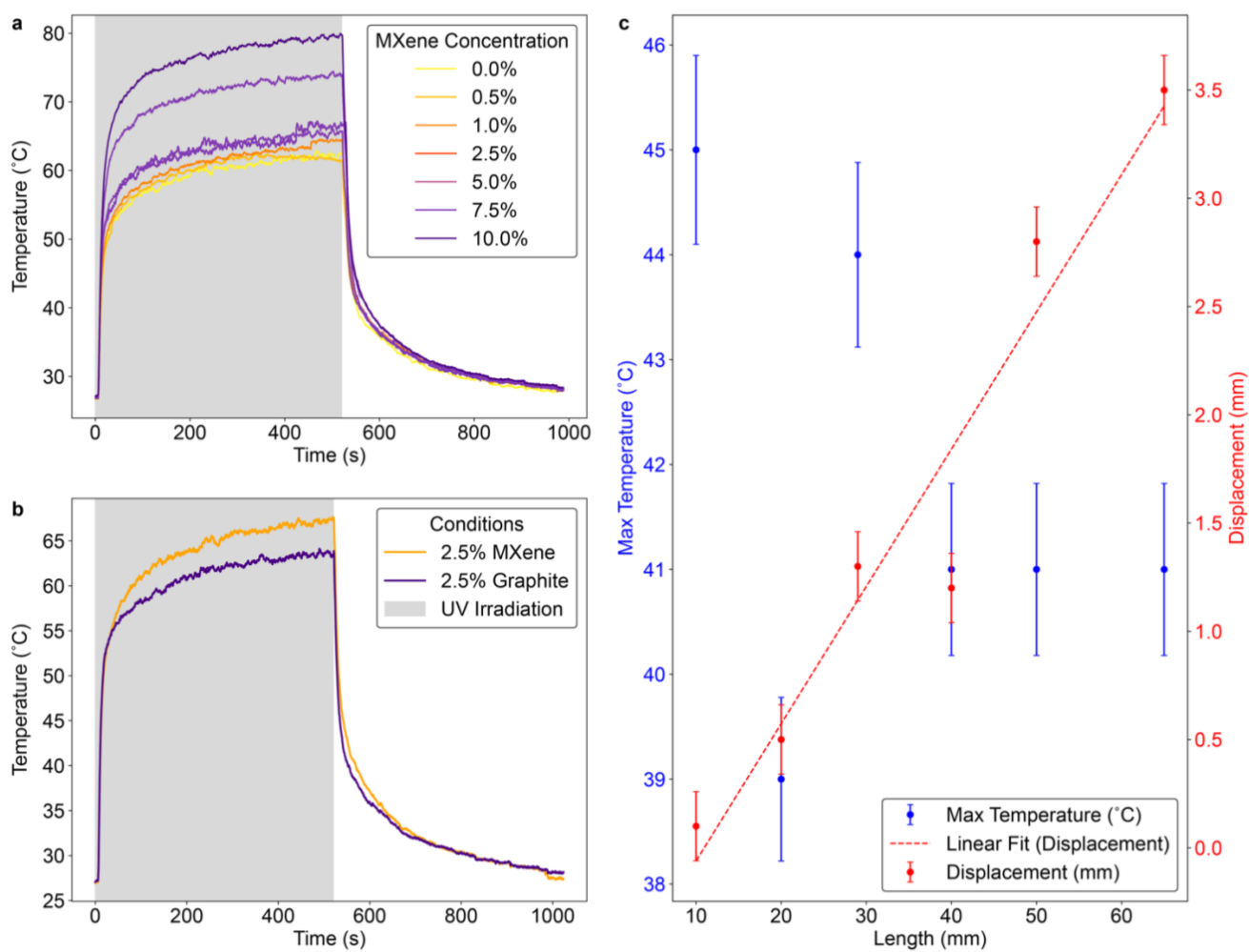

**Fig. 3 | Thermomechanical characterization of the MPTAs. a** Steady-state temperature profiles as a function of the concentration of MXene in the elastomeric layer. The initial 500 s represent the UV irradiation phase, followed by 500 s of thermal dissipation in the dark. **b** Comparison of steady-state temperature profiles between 2.5% MXene MPTA and 2.5% graphite-based trilayer actuator. The MXene trilayer actuator exhibited demonstrably superior thermal conversion efficiency. **c** 2.5% MPTA

temperature and displacement profiles of a 2.5% MPTA as a function of length (10 mm, 20 mm, 29 mm, 40 mm, 50 mm, and 65 mm).

More comprehensively, the thermal responses (Figure 3), irrespective of the MXene concentration, conform to the heating and cooling profiles, as reported previously (36). Specifically, the heating phase is mathematically described by Eq. 3:

$$T = T_{max} + (T_0 - T_{max})\, e^{-(G/mC)\, t} \qquad (3)$$

where $T$, $T_0$, $T_{max}$, $G$, $m$, and $C$ denote the MPTA temperature, ambient temperature, maximum MPTA temperature, heating coefficient, film mass, and heat capacity, respectively. Correspondingly, the cooling phase is described with Eq. 4:

$$T = T_0 + (T_{max} - T_0)\, e^{-(F/mC)\, t} \qquad (4)$$

where $F$ represents the cooling coefficient. Both the heating coefficient, $G$, and the cooling coefficient, $F$ (and consequently the resultant film temperature), exhibit linear dependence on the MXene concentration. Specifically, the fractional increment in the maximum temperature is inversely proportional to the MXene concentration. Therefore, a threshold exists beyond which the incremental material cost associated with increased MXene concentration, in conjunction with the more granular and mechanically fragile microstructure, no longer justifies the marginal increase in the maximum temperature. We conclude that a 2.5% MXene concentration within the MPTA is the optimal selection, balancing cost effectiveness with actuation temperature performance. This concentration was therefore used for subsequent mechanical characterization and other experiments.

Finally, to ensure that the increase in the maximum temperature was not only due to the black color of the MXenes, 2.5% graphite-in-elastomer trilayer actuators were fabricated and tested in the same manner. Figure 3b reveals the enhanced photothermal conversion capabilities of MXenes, which increased the temperature by 4 °C compared to that of their graphitic counterparts and reached a maximum temperature of 63.2 °C. Although the theoretical photothermal conversion efficiency of MXenes is nearly 100%, the improvement achieved by MXenes over graphite was only a few degrees greater than that achieved by MXene particles, unlike in another report (30). This can be partially explained by MXene's light reflectance. The photothermal conversion efficiency of these compounds, which is based on their coating method, is likely to improve in the future.

To demonstrate 2.5% MPTA for small-scale soft robotics applications, the actuation temperature and displacement were tested for different film lengths. The resulting thermal and displacement performances for the 2.5% MPTA are shown in Figure 3c. The maximum temperature recorded for each MPTA varied without a clear trend; the highest temperature was 45 °C for the 10 mm MPTA, while the lowest was 39 °C for the 20 mm MPTA. The 40 mm, 50 mm, and 65 mm MPTA samples all reached similar peak temperatures of approximately 41 °C. Despite these fluctuations, the actuation displacement exhibited a consistent linear increase with increasing MPTA length, with the 10 mm MPTA averaging 0.1 mm of displacement and the 65 mm MPTA reaching 3.5 mm.

*Assessment of MPTA against other photothermomechanical polymer film actuators*

The MPTA performance was defined in terms of four metrics and compared against a few notable examples from three material classes: non-MXene nanoparticles with hydrogels (*54*), non-MXene nanoparticles with liquid crystal elastomers, LCEs (*8*), and MXene-layered polymers in general. These three classes were selected from key literature reviews, where ref. (*17*) quantitatively compares hydrogel and LCE-based actuator performances and ref. (*19*) does the same for MXene-layered polymers. As an overview of the comparison of the four metrics, the MPTA ranked first in three metrics and third in one metric, performing the best overall. For the hydrogel, LCE, and MXene-layered polymer classes, the most representative work performances are provided in Table 1 and will be discussed below. The calculations for each metric are provided in the Supplementary Information.

**Table 1 | Comparison of MPTA to representative photothermomechanical polymer film actuator materials** (the results from this work are in bold and underlined font)

| Materials | Light band (nm) | Irradiation intensity (*) | Bending degrees (†) | Materials cost ($ g$^{-1}$) | Cycles tested (cycles) | Ref. |
|---|---|---|---|---|---|---|
| Non-MXene nanoparticle hydrogel | 450-490 blue | 2.6 | 1.29 | 50 | 10 | (*55*) |
| Non-MXene nanoparticle LCE | 520 cyan | 1.0 | 0.37 | 400 | 10 | (*56*) |
| MXene-layered polymer | 800-25,000 NIR | 6.1 | 0.05 | 5 | 1,000 | (*29*) |
| **MPTA** | **360-370 UV** | **0.1 (1st)** | **0.1 (3rd)** | **2 (1st)** | **1,000 (1st)** | **This work** |

*per °C [mW cm$^{-2}$ °C$^{-1}$]

†per unit irradiation per cycle time, [° mW$^{-1}$ cm$^2$ s$^{-1}$]. Film thicknesses were considered.

*Low irradiation per change in temperature*

Being able to heat a material with minimal energy is fundamental to the efficiency of a photothermomechanical actuator. MPTA requires the least amount of irradiation energy to heat itself, with 0.1 mW cm$^{-2}$ of 365 nm UV irradiation intensity being sufficient to increase its temperature by 1 °C. In comparison, representative non-MXene nanoparticle-embedded hydrogels (*55*), non-MXene nanoparticle-embedded LCEs (*56*), and MXene-layered polymers (*29*) were used at approximately 1 mW cm$^{-2}$ °C$^{-1}$, 2.6 mW cm$^{-2}$ °C$^{-1}$, and 6.1 mW cm$^{-2}$ °C$^{-1}$, respectively. Thus, MPTA can be heated with an irradiation intensity that is an order of magnitude lower than that of other similar materials, minimizing not only energy expenditure but also the risk of high-intensity light in human–robot interactions. These differences could also be attributed to the unique actuation design and architecture of the MPTA. For instance, compared to representative MXene-layered materials (*29*), MPTA uses UV light. In theory, UV light is absorbed by MXenes more efficiently than is NIR radiation, and this phenomenon also holds true for other components in the MPTAs, such as elastomers, plastics, and paper. A material reported by Zuo *et al*. conforms to this conclusion (*56*). We observed that when using a blue light source that corresponds to the peak of the LCE absorbance band, the LCEs require six and two times less irradiation than the respective materials reported in the literature (*29, 55*). Notably, the material in ref. (*55*) also uses a narrow band of absorbance-optimized light. However, this latter work is based on hydrogels, which are actuated by thermally controlled swelling and hygroscopic contraction in an aqueous environment.

*High bending rate*

The simplest metric for evaluating bending performance is the bending angle relative to the tip of the film actuator surface, as described previously (*29, 57*). The film actuators presented here have different thickness-to-length ratios; they are exposed to light of different intensities and cycled over different numbers of cycles, and these factors were considered when calculating the normalized bending degree. The MPTA was calculated to bend $0.1°$ $mW^{-1}$ $cm^2$ $s^{-1}$, making it the third best performing actuator. The typical extent of MPTA motion is demonstrated in Supplementary Movie 1, corresponding to the selected lengths shown in Figure 3c. In comparison, representative non-MXene nanoparticle-embedded hydrogels (*55*), non-MXene nanoparticle-embedded LCEs (*56*), and MXene-layered polymers (*29*) were used at approximately $1.29°$ $mW^{-1}$ $cm^2$ $s^{-1}$, $0.37°$ $mW^{-1}$ $cm^2$ $s^{-1}$, and $0.5°$ $mW^{-1}$ $cm^2$ $s^{-1}$, respectively. This comparison requires a more thorough analysis. MPTA outperforms other MXene-layered polymers owing to its superior MXene photothermal conversion efficiency enabled by its uniform, unagglomerated dispersion and UV-based light input. This response is achieved from the synergetic effect of the thermally expanding elastomer and contracting paper layer. MPTA thus bends more than works that mostly rely on only expanding or contracting the active layer against the passive layer. The Young's moduli of hydrogels reported by Kim *et al*. (*55*) are typically on the order of 1 kPa, up to 1 MPa (*58*), which is more fragile than more polymer/elastomer-based soft materials. Additionally, hydrogels generally require an aqueous environment (*54*), where long-term and reusable anti-dehydration mechanisms become challenging (*59*). An LCE actuator reported by Zuo *et al*. (*56*) requires a high actuation temperature of approximately 70 °C since actuation of LCEs is generally related to heating above their nematic-to-isotropic transition temperature (*15*). Other LCE-based actuators require similar threshold temperatures, heating to approximately 80 °C (*60*), 70 °C (*61*), and 90 °C (*29*). On the other hand, MPTA actuation is not based on such thresholds, allowing it to bend proportionally to temperature differences. This means that it can operate at actuation temperatures as low as 40 to 60 °C and is therefore safer for wearable robotics applications.

*Cost-effectiveness of the design*

In the case of photothermomechanical materials, the bulk of the raw material cost comes from materials such as photothermal nanoparticles, LCEs, and hydrogels, as opposed to more traditional polymers such as polyethylene, polycarbonate, PDMS, and silicone. We thus put particular emphasis on the material cost as one of the metrics for assessing the material for various applications. Non-MXene, nanoparticle-embedded hydrogels (*55*), non-MXene, nanoparticle-embedded LCEs (*56*), and MXene-layered polymers (*29*) currently cost approximately \$50 $g^{-1}$, \$400 $g^{-1}$, and \$5 $g^{-1}$, respectively, suggesting a trade-off between the bending performance and the material cost. Because of its use of low-cost materials, MPTA utilizes MXenes efficiently at a reduced cost; at the current market price, 2.5% of the MPTA costs approximately \$2 $g^{-1}$, making it the least expensive of the three representative actuators above. These considerations are relevant to the projected scaling up of the production of these photothermomechanical polymer film actuators. LCE-based actuators with better performance utilize more expensive precursor materials, such as the mesogenic monomer 4-but-3-enyloxy-benzoic acid 4-methoxy-phenyl ester (MBB) (*56*). The best bending performance of MXene-layered LCE actuators was reported with relatively cheap LCEs based on RM82 and RM257 (*29*). For hydrogel-based actuators, PNIPAm is one of the most commonly used hydrogels (*62*), with a cost of approximately \$100 $g^{-1}$ at higher hydrogel concentrations.

In general, most of the actuators cost between $5–50 g$^{-1}$, as they are based on a combination of a high content of nanoparticles and more traditional polymer materials as host materials.

*Cycle life*

In the assessment of the actuator, we also accounted for the cyclic operation mode, which is the key to the cost and reflects the actuator's durability. Due to the inconsistency of the previously reported data, this analysis required consideration of the number of cycles reported for each material. Representative non-MXene nanoparticle-embedded hydrogels (*55*), non-MXene nanoparticle-embedded LCEs (*56*), and MXene-layered polymers (*29*) were tested for 10 cycles, 10 cycles, and 1,000 cycles, respectively. Compared to these, the MPTA was tested for the largest number of cycles, 1,000 cycles, for more than 60,000 s. Studies outside of these representative works also reported similar magnitudes of cycle life within each material class. While other MXene-layered polymer actuators have been tested for less than 100 cycles, ranging from 1,000 (*31*) to 500 (*57, 63*) and 100 cycles (*64*), other non-MXene nanoparticle-embedded LCE actuators have been tested for fewer than 100 cycles (*56, 60, 61*). Notably, other non-MXene nanoparticle-embedded hydrogel actuators did not exhibit any numbers, capping the manually counted numbers after approximately 10 cycles (*55, 65, 66*). One literature review in soft robotics (*7*) categorizes actuators with "low" durability as those that fail within a few hundred cycles and those with "good" durability as those that can withstand 5,000 cycles. While this does not translate into failure of these actuators in less than a few hundred cycles, it emphasizes that in all cases, extended testing with more cycles is necessary.

*MPTA Application Demonstrations*

As a proof-of-concept of the application of MPTA, we first demonstrate the potential of 2.5% MPTA by using a KIFS design (Figure 4, Supplementary Fig. 1, Supplementary Movie 2). The structure consists of four identically shaped MPTA films stacked into a kirigami-inspired structure resembling an eight-petaled up-side down flower. To demonstrate practical application, the structure was tested by monitoring the temperature and displacements with a common flashlight setup (Supplementary Fig. 2b). During the "light stage," the structure was irradiated for 20 s at 5.4 mW cm$^{-2}$. This was followed by a "dark stage" of 40 s, and the cycles were repeated 1,000 times. The photothermomechanical characterization results in Figures 4a and 4b indicate that over the first 6,000 s, corresponding to 100 cycles, the structure reached an average peak temperature of 49.0 °C during the light stage and cooled to an average minimum temperature of 35.0 °C during the dark stage. The typical increases and decreases in temperature and displacement are plotted from approximately 2,000 s (34$^{th}$ cycle) in Figure 4c. These measurements were taken at the center point, which is the region that experienced the highest temperatures at peak irradiation both at the onset ($t = 0$ s) and peak ($t = 20$ s) of irradiation (Figure 4e and 4e), respectively. Lateral snapshots of the structure showed a 0.78 mm vertical displacement between these two time points, consistent with the laser distance sensor values (Figure 4f and 4g).

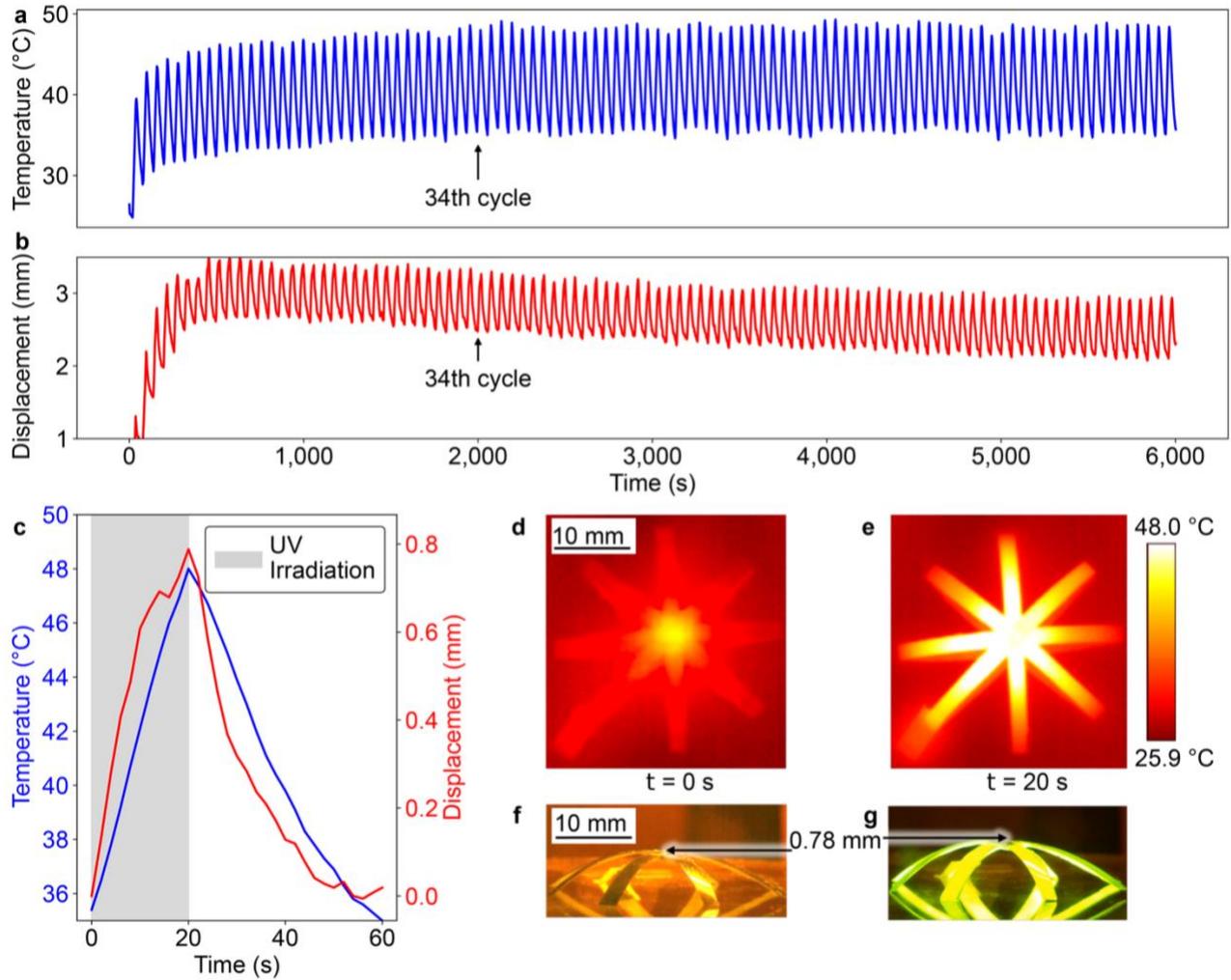

**Fig. 4 | Photothermomechanical characterization of the KIFS MPTA structure. a**, **b** The temperature and displacement (on the same abscissa) are shown over the first 6,000 s of 60,000 s of actuation, corresponding to the first 100 of 1,000 cycles. **c** Zoomed-in representation of the 34[th] cycle from panels a and b. **d**, **e** Thermographic images, recorded from the top, corresponding to the 34[th] cycle before actuation and at maximum displacement. **f**, **g** Lateral views of the actuator corresponding to the points shown in panels d and e.

The overall bending pattern followed a temperature-to-curvature model based on ref. (*31*) and is given by Eq. 5:

$$\varepsilon \propto [(\alpha_e - \alpha_c)\Delta T + \beta_e \Delta C_e - \beta_c \Delta C_c]e^{-k} \quad (5)$$

In Eq. 5, $\varepsilon$, $k$, $\alpha_e$, $\alpha_c$, $\beta_e$, $\beta_c$, $\Delta C_e$, and $\Delta C_c$ are the curvature, plastic photopolymer layer stiffness coefficient, thermal expansion coefficient for the expanding and contracting layers, hygroscopic expansion coefficients, and variation in moisture concentration, respectively. During the first 1800 s (30 cycles), the maximum temperature increased with each cycle, although the temperature increase per cycle remained relatively constant, averaging 12.7 °C per cycle. After this stabilization period, the change in temperature became steady, with the system reaching an average maximum temperature of 48.9 °C in the light stage and an average minimum temperature of 35.0 °C in the dark stage. This 14.9 °C average temperature increase per cycle is lower than the 19.1 °C average increase observed in the 100-cycle test and translates into reduced actuation capability. In contrast, the average displacement decreased by approximately 0.8 mm over time. After 10,200 s, equivalent to 170 cycles, the actuation stabilized, averaging 0.8 mm displacement per cycle, which is 0.02 mm larger than that of the 100-cycle test snapshot.

Furthermore, the flower MPTA characterization included 5,000 s light and dark stage steady-state tests. The same procedure was used for the images shown in Figure 3a and 3b (Supplementary Fig. 2a). As illustrated in Figure S4c, the actuator reached a temperature of 66.3 °C, and the average stable actuation displacement was 0.94 mm after approximately 130 seconds. The mixture was heated for an additional 300 s without significant changes in the actuation displacement. The temperature and displacement peaked at 70.2 °C and 0.91 mm, respectively. Once the irradiation was turned off, the actuator was allowed to cool and unbend. The consistent temperature and displacement performance seem to be enabled by robust interfacial contact between the three MPTA layers. Supplementary Figure 4d shows that the legs of the flower MPTA before and after 1,000 cycles of actuation do not exhibit delamination. We hypothesize, based on the chemical structure and composition of the components, that the elastomer and plastic photopolymer layers adhere to each other through covalent bonds, while the plastic and paper layers are joined by both covalent bonds and interlocking of the cellulose fibers. The consistent displacement observed during the 1,000-cycle test highlights the high cycle life of the actuator over an extended period of operation.

Aside from its decorative potential like real flowers, the flower MPTA's performance was further characterized. We measured the force exerted by the MPTA over 6,000 s (100 cycles; Supplementary Fig. 4e). The maximum forces, averaging 7.9 mN, align with the peak irradiation and occur just before the UV light is turned off. To evaluate the vertical stiffness, the vertical force from the 100-cycle test is plotted in Supplementary Figure 4e against the displacement data from the final 6,000 s (100 cycles) of the 60,000 s (1,000 cycles) test in Supplementary Figure 4b, considering only the light stage of each cycle. In Supplementary Figure 4f, the resulting trend line reveals a stiffness of 11.4 N m$^{-1}$. As discussed above (Eq. 5), the lower the stiffness of the actuator is, the easier it is to bend. In contrast, the actuator exerts more force with higher stiffness. Since the stiffness is based on the springy curved topology of the MPTA, Young's modulus may be a better indicator of its innate durability. For reference, three 5 × 31 mm MXene-trilayer samples (0%, 2.5%, 5%) and a paper sample were subjected to tensile testing to determine their elastic modulus. We reported an average experimental elastic modulus of 2.93 ± 0.47 GPa (Supplementary Fig. 5).

We utilized MPTA's stiffness and force to show a second demonstration in Fig. 5a-d and Supplementary Movie 3. In Fig 5a,b, a parallel manipulator with hexagonal paper platform is held up by three 10 × 20 mm 2.5% MPTAs. When irradiated for 100 s, the structure lowers the platform by 0.96 mm, and reverted after 100 s of no irradiation. In Fig 5c,d, We show that individual film can be actuated, controlling the platform's orientation. For visibility, the setup from Fig 5a,b is simplified to two films, with only the left side being irradiated. When irradiated for 100 s, the structure tilted the platform leftward by 5°, and reverted after 100 s of no irradiation. Variable height and orientation platforms of this size, scalable to large array, are useful for camouflaging (*67, 68*), antenna control (*69*), and acoustic (*70*) and optical waveguides (*24*), as these works function under vis-NIR-IR and could benefit from adding UV-controlled degrees of freedom.

As a third demonstration, we show in Fig. 5e-j that MPTAs' stiffness and force exertion can also be utilized as a soft gripper for soft manufacturing (*71*), specifically to decorate a marshmallow on a chocolate muffin without any tethered actuation source. Its setup is detailed in Supplementary Fig. 6 and played out in Supplementary Movie 4. To move the gripper, we use a magnetically controlled arm to position the gripper from Fig. 5e to f in 11 s. We show in Fig. 5g to h that once the UV light is turned on for around 80 s, the gripper has enough force to live the marshmallow, and move it above the muffin in 30 s. Once the light is turned off in Fig. 5i, the marshmallow is released and laid flat on the muffin after 60 s. Such a setup is beneficial over conventional motor-based cake assembly, as it not only handles delicate samples, but also stay free of degradation and malfunctioning due to kitchen powder, vapor, and liquid accumulation. Its low cost nature of the setup is ideal for automating constantly renewed menu items, especially seasonally renewed airline foods that are handmade to this date (*72*). The benefit of using specifically UVA for soft gripper as opposed to other wavelength such as vis or IR is that these can respectively be used instead for computer vision-based environment sensing to adapt the gripper and the arm's control real-time. The exact gripper sensing implementation is left for future work.

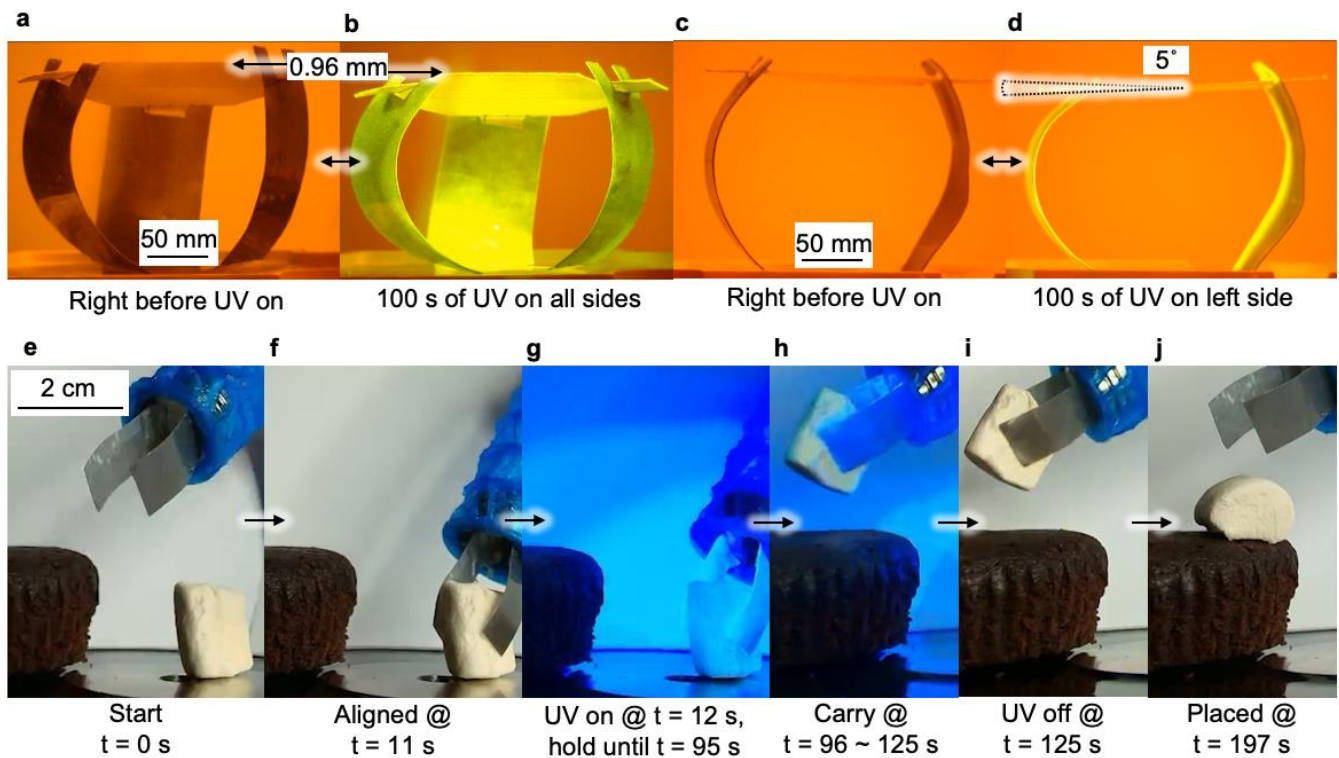

**Fig. 5 | Utilization of MPTA stiffness and force for parallel manipulator and soft gripper. a,b** Resting and fully photo-actuated manipulator platform height controlled by three 10 × 20 mm 2.5% MPTA supports. **c,d** Manipulator platform orientation controlled by photo-actuating one of the two 10 × 20 mm 2.5% MPTA support. **e-j** UVA-irradiated soft gripper for assembling marshmallow on top of a chocolate muffin.

## Discussion

This study describes the design and performance of a novel UV-driven photothermomechanical actuator (MPTA) with an MXene-dispersed trilayer structure. MPTA exhibits high UV-to-heat conversion efficiency coupled with strong interfacial adhesion between the constituent layers and effective photothermal dispersion of MXene within the elastomeric layer. Thermomechanical analysis revealed that the optimal composition of 2.5% MPTA surpasses both MXene-based and non-MXene-based polymeric actuators in terms of efficiency, cost effectiveness, and durability upon cycling. MPTA application prototypes demonstrated sustained bending and force exertion for 1,000 cycles over 60,000 s, parallel manipulator, and soft gripper. The results suggest that MPTAs offer a viable solution to some of the challenges in soft robotics (*73*), including complex robotic gripping, locomotion, and manipulation, as well as easy-to-access rapid development for simpler designs (*74*). Furthermore, the light-driven actuation capability of MPTA is suitable for applications beyond robotics, such as haptics, thereby addressing the grand challenges of soft robotics (*1*).

Future research on these materials will aim to enhance the photothermal conversion efficiency through MXene optimization, functionalization, and tailoring of material properties. Admittedly, UV-vis light are known to degrade polymers (*75*) and harm bio tissues in the long-term (*76*), and implementing sustainable anti-degradation and safety mechanisms for deploying the above applications were left out of scope. Improved characterization of photothermomechanical processes, especially performance under full UVA-to-IR and more comprehensive fatigue analysis on the host polymers, are essential. While current assessments focus on bending, future studies should evaluate the force, work, and efficiency of these materials given the high Young's modulus and potential for superior performance of the MPTA, and do so through more automated fabrication to minimize sample variance.

## Materials and Methods

### *Materials*

In the selection of elastomers, while LCEs offer anisotropic expansion, their high cost compared to that of isotropic elastomers necessitates the consideration of alternative materials. A commercially available elastomer resin (Resione F80 Elastic 3D Printer Resin Black) was chosen for the expanding layer, providing a balance between thermal expansion characteristics and cost. The elastomer has a high content of acrylated aliphatic urethane, a composition known to enhance the compliance and thermal expansion properties relative to other elastomeric materials (*77*). As a thermosetting photopolymer, this resin facilitates homogeneous dispersion of MXene nanoparticles and subsequent facile curing by exposure to UV radiation. For the fabrication of the intermediate layer, a plastic resin (Esun Hard-Tough Resin) was selected. The high urethane acrylate content of this resin confers enhanced structural integrity, hardness, adhesion, and chemical resistance, facilitating the formation of a cohesive trilayer architecture. Consistent with the elastomeric layer, this plastic resin is a thermosetting photopolymer that undergoes rapid polymerization upon exposure to UV irradiation. Notably, the urethane acrylate components of the plastic resin form a cohesive interface with the acrylated aliphatic urethane of the elastomer top layer. This observation is supported by preliminary studies that noted a significantly greater incidence of delamination during MPTA fabrication in the absence of an intermediate plastic layer (*45*). The choice of the bottom layer was set to 75 GSM paper substrates, following an evaluation of papers with varying weights (*45*).

The 75 GSM paper substrate is characterized by a thickness of approximately 100 μm, consistent with standard office paper specifications.

*Fabrication of the MPTA*

The fabrication of MPTA, summarized in Supplementary Fig. 1, starts with sieving MXene particulates through a 200-mesh screen with a 75-micron aperture (Supplementary Fig. 1b). The sieved particulates are subsequently incorporated into the elastomer resin. The concentrations of MXene relative to the elastomer were 0.0%, 0.5%, 1.0%, 2.5%, 5.0%, 7.5%, and 10.0%. To achieve a homogeneous dispersion, the elastomer-MXene composite was subjected to grinding with an agate mortar and pestle. Agate was selected due to its chemical inertness, high stiffness, and lack of porosity, which mitigates contamination and unwanted reactions, and due to its smooth surface, which minimizes material attrition. After grinding, the MXene suspension was degassed under vacuum for 10 min to eliminate trapped air bubbles, thus preventing the formation of voids that could compromise the homogeneity of the mixture and adversely affect the photothermal response of the actuator (Supplementary Fig. 1c). The degassed suspension was then dispensed onto a glass wafer (Supplementary Fig. 1e) and spun at 1,000 rpm for 30 s using a spin-coater Laurell WS-650 (Supplementary Fig. 1f). The selection of the coating thickness was based on a previous study (*45*). Finally, as depicted in Supplementary Figure 1g, the coated layer was cured by exposure to 365 nm UV radiation. Given that the elastomer was premixed with photoinitiators, the layer was transformed into the photothermally expanding layer of the MPTA within 30 s. After curing, the trilayer laminate (Supplementary Fig. 1i) was detached from the glass wafer via a peeling process performed in a direction orthogonal to the fiber alignment (Supplementary Fig. 1j). A rigid rod was employed to facilitate detachment, ensuring uniform and controlled release of the film while mitigating potential damage. The resultant film was then sectioned into strips of varying aspect ratios, tailored to the requirements of subsequent characterization procedures. For each MXene concentration, the longitudinal axis of the strips was consistently oriented orthogonal to the fiber alignment for each concentration (Supplementary Fig. 1k).

The overall fabrication process is characterized by low material costs and procedural simplicity. A quantity of 10 g of MXene (80–100 nm thick, 99% purity $Ti_3C_2T_x$) from Nanochemazone was procured at a cost of approximately $650. While the high purity of MXenes facilitates efficient photothermal conversion, the associated cost necessitates the development of actuators utilizing low MXene concentrations and cost-effective host materials. Since 1,000 g of the elastomer from Resione and the plastic photopolymers from Esun cost approximately $110 and $140, respectively, the fabrication of MPTA with MXene-elastomer concentrations ranging from 0% to 10% incurs a cost between $0.1 $g^{-1}$ and $6 $g^{-1}$ (assuming minimal loss during fabrication), wherein MXene comprises between 83–99% of the cost, respectively (Supporting Information). This cost effectiveness is substantiated by the comparative data presented in Table 1, which demonstrates favorable comparisons with other photothermomechanical film actuators. Furthermore, the fabrication of an MPTA on a 10 cm glass wafer was completed within a timeframe of approximately two hours.

*Characterization Setup*

*MPTA*

Photothermal characterization was conducted by using electron microscopy, spectroscopy, and film thickness measurements. For electron microscopy, a ThermoFischer Quanta 450 FEG scanning electron microscope (SEM) was used. MPTA films with dimensions of 2 × 4 mm were subjected to gold coating using a Cressington 108 Auto Sputter Coater for SEM imaging. Elemental mapping was performed using EDAX APEX Software on a Z4-i7+ Analyzer from AMETEK. The surface area of the MXene nanosheets within the obtained image was quantified using ImageJ software. To determine the MPTA layer thickness, the thickness after fabrication was measured using a Bruker Dektek XT-A surface profiler. Thickness measurements were acquired from the glass wafer base to the coated MPTA layer prior to detachment. For spectroscopic analysis, an INVENIO-S Fourier Transform Infrared Spectrometer, a Jasco FP-8500 Spectrofluorometer, and a Shimadzu UV-Visible Spectrophotometer UV-2600i were utilized. MPTA samples with a size of 25 × 75 mm were prepared for these analyses. To quantify the temperature response of the MPTA films under each set of conditions, a thermography apparatus was designed and used, as depicted in Supplementary Fig. 2a. The apparatus included a 65 W 365 nm UV light source, which provided an irradiation intensity of approximately 25 mW cm$^{-2}$ for a duration of 500 s, followed by a 500 s period without irradiation to facilitate the acquisition of converging readings. The irradiation intensity was determined based on measurements from Torch Bearer's Y2 spectrometer, 3.5~cm away from the flashlight. The UV source was positioned upward in direct contact with a glass wafer on which 16 MPTA films, each measuring 5 × 10 mm and representing three replicates per MXene concentration, were placed. The perimeter of the experimental setup was enclosed with UV filter panels to prevent extraneous UV exposure. The perimeter of the experimental setup was enclosed with UV filter panels to prevent extraneous UV exposure. Temperature profiles for each film were simultaneously recorded over time using a Gobi-640 thermal camera from Exosens©. This camera exhibited a temperature measurement range from approximately 20 °C to 80 °C with an accuracy of approximately 2%. A thermal camera was mounted above the film sample to capture the irradiated surface area. Temperature data were extracted from the central pixel of each film's thermal image to document temperature variations over time.

Considering the importance of demonstrating MPTA for small-scale soft robotics applications, ideal film lengths were tested via temperature, displacement, and video recording. For this purpose, a modified version of the setup in Supplementary Fig. 2a was used. This is shown in Supplementary Fig. 2b, where 3 mm wide MPTA films of varying length and concentration were tested one at a time. As a preliminary study showed that width did not affect the performance of a 10 mm long trilayer actuator, the tested film width was kept at 3 mm, which balanced compactness and stable mounting for measurement purposes (*45*). While the actuation performance also stayed constant (within an approximately 0.1 mm difference) for different layer thicknesses, those that achieved the highest actuation were chosen (*45*) (0.03 mm elastomer photopolymer, 0.05 mm plastic photopolymer, 0.1 mm paper). Notably, once MXene was added to the elastomeric layer, the thickness increased to 0.07 mm. The 3 mm wide 2.5% MPTA with various lengths (10 mm, 20 mm, 29 mm, 40 mm, 50 mm, and 65 mm) each underwent 10 cycles of testing, with each cycle consisting of a 20-second UV irradiation phase followed by a 40 s cooling period. A uvBeast™ V3 365 nm filtered black light/UV flashlight, which is rated at approximately 5.4 mW cm$^{−2}$, was used. . This was confirmed with Torch Bearer's Y2 spectrometer, measuring 3.5~cm away from the flashlight, which gave 5.9 mW cm$^{−2}$ nm$^{-1}$ with 15 nm full width half maximum (FWHM) reading at 365 nm (100 mW/cm² when integrating between 340 to 402~nm spectrum). The flashlight was facing upward and in contact with a glass wafer upon which the MPTA was placed.

For displacement recording, an optoNCDT 1420-200 optical noncontact displacement sensor was chosen for the high sensing linearity of ≤ ±0.16 mm. The sensor laser was aligned to point to the center of the film to capture peak displacement. Finally, a video recording was performed using a Dino-Lite Digital Microscope. The microscope was positioned behind a UV filter panel to reduce UV exposure. The displacement observed from the recorded video, analyzed through Tracker® software, was compared with the optical displacement values to cross-check the data.

*KIFS MPTA*

The KIFS MPTA design was based on Tabassian *et al.*'s work on photothermomechanical kirigami flower actuators (*14*), which also use a photothermally contracting cellulose layer. Due to the size constraints of the tensile tester used, a length of 29 mm was selected, and the 3 mm width was decreased to 2 mm to help the MPTA bend together. Specifically, four sets of 2 × 29 mm films were cut and stacked on top of each other and bonded at the midpoint.

We used the Fullam SEMTester® manufactured by MTI Instruments (Supplementary Fig. 3), which was equipped with a 5 N load cell that provides a sensing accuracy of ±0.2%. The tester was positioned vertically and fitted with an aluminum base to hold the KIFS MPTA. The load cell was fitted with a protruding fixture. Before each measurement, the base height was adjusted so that the KIFS MPTA's center point slightly contacted the fixture. At this point, the recording software's reading was calibrated to zero. The KIFS MPTA was also irradiated with the same 5.4 mW cm$^{-2}$ UV light pattern, positioned 1.5 cm away at a 30° incident angle. The flashlight was turned on and off with a transistor switch controlled by an Atmega328p microcontroller unit (MCU).

*Parallel Manipulator*

The three 10 × 20 mm 2.5% MPTAs and the 10 mm radius hexagonal paper platform for Fig. 5a,b, or the 12 × 20 mm rectangular paper platform for Fig. 5c,d, are connected together like tongue-and-groove-like kirigami joints. Each of the MPTAs' top end has a 4 × 5 mm groove cut out, while the correspond end of the platforms are cut out to have 4 × 5 mm. Such a joint keeps the structure together with enough friction under minimal bending constraints. For anchoring, the MPTAs were taped on a base. The tapes act as compliant hinges to constrain the MPTA from bending sideways.

*Soft Gripper Specifications*

Two 10 × 20 mm 2.5% MPTA were hot glued on triple-stacked 20 × 3 mm neodymium coin magnets as seen in Fig. 5e, each with 47.7 mT. The magnets were fitted on the tip of a 3D printed thermoplastic polyurethane (TPU) arm. The soft gripper is vertically steered using a 59.0 mT electromagnet connected to ATmega328P-controlled H-bridge and 11V, 1A DC power supply. The gripper is horizontally steered with manually held 200 mT magnet. The magnetic flux densities were measured on the surface of the magnets using the PS-3221 magnetic field sensor from PASCO. The arm's Kresling origami-inspired spring structure is a more durable, single-material, perforated variant of our previous work (*78*).

**Data and materials availability**

All the data are available from the corresponding authors upon request. Source data are provided with this paper.

# References

1. G.-Z. Yang, J. Bellingham, P. E. Dupont, P. Fischer, L. Floridi, R. Full, N. Jacobstein, V. Kumar, M. McNutt, R. Merrifield, The grand challenges of science robotics. *Science robotics* **3**, eaar7650 (2018).
2. G. Bao, H. Fang, L. Chen, Y. Wan, F. Xu, Q. Yang, L. Zhang, Soft robotics: Academic insights and perspectives through bibliometric analysis. *Soft robotics* **5**, 229--241 (2018).
3. Y. Zhou, H. Li, A scientometric review of soft robotics: Intellectual structures and emerging trends analysis (2010--2021). *Frontiers in Robotics and AI* **9**, 868682 (2022).
4. C. Majidi, Soft-matter engineering for soft robotics. *Advanced Materials Technologies* **4**, 1800477 (2019).
5. Y. Jung, K. Kwon, J. Lee, S. H. Ko, Untethered soft actuators for soft standalone robotics. *Nature Communications* **15**, 3510 (2024).
6. J. Qu, Z. Yu, W. Tang, Y. Xu, B. Mao, K. Zhou, Advanced technologies and applications of robotic soft grippers. *Advanced Materials Technologies* **9**, 2301004 (2024).
7. L. Hines, K. Petersen, G. Z. Lum, M. Sitti, Soft actuators for small-scale robotics. *Advanced materials* **29**, 1603483 (2017).
8. O. Yasa, Y. Toshimitsu, M. Y. Michelis, L. S. Jones, M. Filippi, T. Buchner, R. K. Katzschmann, An overview of soft robotics. *Annual Review of Control, Robotics, and Autonomous Systems* **6**, 1--29 (2023).
9. Y. Zhao, M. Hua, Y. Yan, S. Wu, Y. Alsaid, X. He, Stimuli-responsive polymers for soft robotics. *Annual Review of Control, Robotics, and Autonomous Systems* **5**, 515--545 (2022).
10. Y. Xiong, L. Zhang, P. Weis, P. Naumov, S. Wu, A solar actuator based on hydrogen-bonded azopolymers for electricity generation. *Journal of Materials Chemistry A* **6**, 3361-3366 (2018).
11. S. C. Sahoo, N. K. Nath, L. Zhang, M. H. Semreen, T. H. Al-Tel, P. Naumov, Actuation based on thermo/photosalient effect: a biogenic smart hybrid driven by light and heat. *RSC advances* **4**, 7640-7647 (2014).
12. H. Zeng, P. Wasylczyk, D. S. Wiersma, A. Priimagi, Light robots: bridging the gap between microrobotics and photomechanics in soft materials. *Advanced Materials* **30**, 1703554 (2018).
13. L. Zhang, I. Desta, P. Naumov, Synergistic action of thermoresponsive and hygroresponsive elements elicits rapid and directional response of a bilayer actuator. *Chemical Communications* **52**, 5920-5923 (2016).
14. R. Tabassian, M. Mahato, S. Nam, V. H. Nguyen, A. Rajabi-Abhari, I.-K. Oh, Electro-Active and Photo-Active Vanadium Oxide Nanowire Thermo-Hygroscopic Actuators for Kirigami Pop-up. *Advanced Science* **8**, 2102064 (2021).
15. J. Lagerwall, Liquid crystal elastomer actuators and sensors: Glimpses of the past, the present and perhaps the future. *Programmable Materials* **1**, e9 (2023).
16. Y. Ma, Y. Zhang, L. Yang, H. Qin, W. Liang, C. Zhang, Light-Driven Small-Scale Soft Robots: Material, Design and Control. *Smart Materials and Structures*,  (2024).
17. Y. Chen, J. Yang, X. Zhang, Y. Feng, H. Zeng, L. Wang, W. Feng, Light-driven bimorph soft actuators: design, fabrication, and properties. *Materials Horizons* **8**, 728--757 (2021).
18. Y. Yang, Y. Shen, Light-Driven Carbon-Based Soft Materials: Principle, Robotization, and Application. *Advanced Optical Materials* **9**, 2100035 (2021).
19. J. He, P. Huang, B. Li, Y. Xing, Z. Wu, T.-C. Lee, L. Liu, Untethered Soft Robots Based on 1D and 2D Nanomaterials. *Advanced Materials*, 2413648 (2025).

20. M. W. Barsoum, T. El-Raghy, Synthesis and characterization of a remarkable ceramic: Ti3SiC2. *Journal of the American Ceramic Society* **79**, 1953--1956 (1996).
21. M. Shekhirev, C. E. Shuck, A. Sarycheva, Y. Gogotsi, Characterization of MXenes at every step, from their precursors to single flakes and assembled films. *Progress in Materials Science* **120**, 100757 (2021).
22. Y. Wang, T. Guo, Z. Tian, L. Shi, S. C. Barman, H. N. Alshareef, MXenes for soft robotics. *Matter* **6**, 2807--2833 (2023).
23. B. Anasori, Y. Gogotsi, MXenes: trends, growth, and future directions. *Graphene and 2D Materials* **7**, 75--79 (2022).
24. X. Yang, L. Lan, L. Li, J. Yu, X. Liu, Y. Tao, Q.-H. Yang, P. Naumov, H. Zhang, Collective photothermal bending of flexible organic crystals modified with MXene-polymer multilayers as optical waveguide arrays. *Nature communications* **14**, 3627 (2023).
25. D. Xu, Z. Li, L. Li, J. Wang, Insights into the photothermal conversion of 2D MXene nanomaterials: synthesis, mechanism, and applications. *Advanced Functional Materials* **30**, 2000712 (2020).
26. Y. Li, C. Xiong, H. Huang, X. Peng, D. Mei, M. Li, G. Liu, M. Wu, T. Zhao, B. Huang, 2D Ti3C2Tx MXenes: Visible Black but Infrared White Materials. *Advanced Materials* **33**, 2103054 (2021).
27. M. Yang, Y. Xu, X. Zhang, H. K. Bisoyi, P. Xue, Y. Yang, X. Yang, C. Valenzuela, Y. Chen, L. Wang, Bioinspired phototropic MXene-reinforced soft tubular actuators for omnidirectional light-tracking and adaptive photovoltaics. *Advanced Functional Materials* **32**, 2201884 (2022).
28. X. Yang, Y. Chen, F. Wang, S. Chen, Z. Cao, Y. Feng, L. Wang, W. Feng, Bioinspired artificial phototaxis and phototropism enabled by photoresponsive smart materials. *Materials Today*,  (2025).
29. W. Cho, D. J. Kang, M. J. Hahm, J. Jeon, D.-G. Kim, Y. S. Kim, T. H. Han, J. J. Wie, Multi-functional locomotion of collectively assembled shape-reconfigurable electronics. *Nano Energy* **118**, 108953 (2023).
30. H. Shin, W. Jeong, T. H. Han, Maximizing light-to-heat conversion of Ti3C2Tx MXene metamaterials with wrinkled surfaces for artificial actuators. *Nature Communications* **15**, 1--11 (2024).
31. G. Cai, J.-H. Ciou, Y. Liu, Y. Jiang, P. S. Lee, Leaf-inspired multiresponsive MXene-based actuator for programmable smart devices. *Science advances* **5**, eaaw7956 (2019).
32. J. Li, R. Zhang, L. Mou, M. Jung de Andrade, X. Hu, K. Yu, J. Sun, T. Jia, Y. Dou, H. Chen, F. Shaoli, Q. Dong, L. Zunfeng, Photothermal bimorph actuators with in-built cooler for light mills, frequency switches, and soft robots. *Advanced Functional Materials* **29**, 1808995 (2019).
33. J. Li, L. Mou, R. Zhang, J. Sun, R. Wang, B. An, H. Chen, K. Inoue, R. Ovalle-Robles, Z. Liu, Multi-responsive and multi-motion bimorph actuator based on super-aligned carbon nanotube sheets. *Carbon* **148**, 487--495 (2019).
34. S. Jeon, C. Lee, D. K. Yoon, Dual Stimuli-Responsive Actuation of Oriented MXene/Polymer Composite. *ACS Applied Materials & Interfaces*,  (2025).
35. S. Ma, P. Xue, Y. Tang, R. Bi, X. Xu, L. Wang, Q. Li, Responsive soft actuators with MXene nanomaterials. *Responsive Materials*, e20230026 (2024).
36. R. Li, L. Zhang, L. Shi, P. Wang, MXene Ti3C2: an effective 2D light-to-heat conversion material. *ACS nano* **11**, 3752--3759 (2017).

37. A. Lipatov, A. Goad, M. J. Loes, N. S. Vorobeva, J. Abourahma, Y. Gogotsi, A. Sinitskii, High electrical conductivity and breakdown current density of individual monolayer Ti3C2Tx MXene flakes. *Matter* **4**, 1413-1427 (2021).
38. N. S. Vorobeva, S. Bagheri, A. Torres, A. Sinitskii, Negative photoresponse in Ti3C2T x MXene monolayers. *Nanophotonics* **11**, 3953-3960 (2022).
39. S. Chertopalov, V. N. Mochalin, Environment-sensitive photoresponse of spontaneously partially oxidized Ti3C2 MXene thin films. *ACS nano* **12**, 6109-6116 (2018).
40. V. Bacheva, A. Firouzeh, E. Leroy, A. Balciunaite, D. Davila, I. Gabay, F. Paratore, M. Bercovici, H. Shea, G. Kaigala, Dynamic control of high-voltage actuator arrays by light-pattern projection on photoconductive switches. *Microsystems & Nanoengineering* **9**, 59 (2023).
41. Z. Ma, H. Joh, D. E. Fan, P. Fischer, Dynamic ultrasound controlled by light. *Advanced Science* **9**, 2104401 (2022).
42. H. Li, J. Dai, Z. Wang, H. Zheng, W. Li, M. Wang, F. Cheng, Digital light processing (DLP)-based (bio) printing strategies for tissue modeling and regeneration. *Aggregate* **4**, e270 (2023).
43. Z. Jiang, M. R. M. Atalla, G. You, L. Wang, X. Li, J. Liu, A. M. Elahi, L. Wei, J. Xu, Monolithic integration of nitride light emitting diodes and photodetectors for bi-directional optical communication. *Opt. Lett.* **39**, 5657-5660 (2014).
44. P. Zhang, L. T. de Haan, M. G. Debije, A. P. Schenning, Liquid crystal-based structural color actuators. *Light: Science & Applications* **11**, 248 (2022).
45. K. Iiyoshi, G. Korres, O. Nagy, G. Roldán, M. Eid, in *2025 IEEE 8th International Conference on Soft Robotics (RoboSoft)*. (IEEE, 2025), pp. 1-6.
46. M. F. Ashby, *Materials Selection in Mechanical Design* (Butterworth-Heinemann, ed. 5, 2016).
47. M. Amjadi, M. Sitti, High-performance multiresponsive paper actuators. *ACS nano* **10**, 10202--10210 (2016).
48. W. Yan Liu and Siyao Shang and Shuting Mo and Peng Wang and Bin Yin and Jiaming, Soft actuators built from cellulose paper: A review on actuation, material, fabrication, and applications. *Journal of Science: Advanced Materials and Devices* **6**, 321-337 (2021).
49. G. Tang, X. Zhao, S. Liu, D. Mei, C. Zhao, L. Li, Y. Wang, Moisture-Driven Actuators. *Advanced Functional Materials* **35**, 2412254 (2025).
50. T. Parker, D. Zhang, D. Bugallo, K. Shevchuk, M. Downes, G. Valurouthu, A. Inman, B. Chacon, T. Zhang, C. E. Shuck, Y.-J. Hu, Y. Gogotsi, Fourier-Transform Infrared Spectral Library of MXenes. *Chemistry of materials*, (2024).
51. C. Zhou, Z. Li, S. Liu, L. Ma, T. Zhan, J. Wang, Synthesis of MXene-based self-dispersing additives for enhanced tribological properties. *Tribology Letters* **70**, 63 (2022).
52. X. Zhu, P. Jiang, Y. Leng, M. Lu, X. Mei, J. Shen, Y. Fan, H. Lu, Synthesis and properties of an eco-friendly bio-based plasticizer for PVC with enhanced migration resistance. *Journal of Materials Science* **59**, 17730--17746 (2024).
53. X. Cui, Q. Ruan, X. Zhuo, X. Xia, J. Hu, R. Fu, Y. Li, J. Wang, H. Xu, Photothermal Nanomaterials: A Powerful Light-to-Heat Converter. *Chemical Reviews* **123**, 6891-6952 (2023).
54. Y. Gao, X. Wang, Y. Chen, Light-driven soft microrobots based on hydrogels and LCEs: development and prospects. *RSC advances* **14**, 14278--14288 (2024).
55. D. Kim, H. S. Lee, J. Yoon, Highly bendable bilayer-type photo-actuators comprising of reduced graphene oxide dispersed in hydrogels. *Scientific reports* **6**, 20921 (2016).
56. B. Zuo, M. Wang, B.-P. Lin, H. Yang, Visible and infrared three-wavelength modulated multi-directional actuators. *Nature communications* **10**, 4539 (2019).

57. Y. Hu, L. Yang, Q. Yan, Q. Ji, L. Chang, C. Zhang, J. Yan, R. Wang, L. Zhang, G. Wu, J. Sun, B. Zi, W. Chen, Y. Wu, Self-locomotive soft actuator based on asymmetric microstructural ti3c2t x mxene film driven by natural sunlight fluctuation. *ACS nano* **15**, 5294--5306 (2021).
58. Z. Li, J. Lu, T. Ji, Y. Xue, L. Zhao, K. Zhao, B. Jia, B. Wang, J. Wang, S. Zhang, Z. Jiang, Self-Healing Hydrogel Bioelectronics. *Advanced Materials* **36**, 2306350 (2024).
59. Y. Chen, Y. Zhang, H. Li, J. Shen, F. Zhang, J. He, J. Lin, B. Wang, S. Niu, Z. Han, Z. Guo, Bioinspired hydrogel actuator for soft robotics: Opportunity and challenges. *Nano Today* **49**, 101764 (2023).
60. R. R. Kohlmeyer, J. Chen, Wavelength-selective, IR light-driven hinges based on liquid crystalline elastomer composites. *Angewandte Chemie International Edition* **52**, (2013).
61. S. Banisadr, J. Chen, Infrared actuation-induced simultaneous reconfiguration of surface color and morphology for soft robotics. *Scientific reports* **7**, 17521 (2017).
62. J. Liu, L. Jiang, S. He, J. Zhang, W. Shao, Recent progress in PNIPAM-based multi-responsive actuators: A mini-review. *Chemical Engineering Journal* **433**, 133496 (2022).
63. M. Xu, L. Li, W. Zhang, Z. Ren, J. Liu, C. Qiu, L. Chang, Y. Hu, Y. Wu, MXene-Based Soft Actuators with Multiresponse and Diverse Applications by a Simple Method. *Macromolecular Materials and Engineering* **308**, 2300200 (2023).
64. L. Xu, H. Zheng, F. Xue, Q. Ji, C. Qiu, Q. Yan, R. Ding, X. Zhao, Y. Hu, Q. Peng, X. He, Bioinspired multi-stimulus responsive MXene-based soft actuator with self-sensing function and various biomimetic locomotion. *Chemical Engineering Journal* **463**, 142392 (2023).
65. Q. Shi, H. Xia, P. Li, Y.-S. Wang, L. Wang, S.-X. Li, G. Wang, C. Lv, L.-G. Niu, H.-B. Sun, Photothermal Surface Plasmon Resonance and Interband Transition-Enhanced Nanocomposite Hydrogel Actuators with Hand-Like Dynamic Manipulation. *Advanced Optical Materials* **5**, 1700442 (2017).
66. Z. Chen, R. Cao, S. Ye, Y. Ge, Y. Tu, X. Yang, Graphene oxide/poly (N-isopropylacrylamide) hybrid film-based near-infrared light-driven bilayer actuators with shape memory effect. *Sensors and Actuators B: Chemical* **255**, 2971--2978 (2018).
67. Y. Liu, R. Bi, X. Zhang, Y. Chen, C. Valenzuela, Y. Yang, H. Liu, L. Yang, L. Wang, W. Feng, Cephalopod-Inspired MXene-Integrated Mechanochromic Cholesteric Liquid Crystal Elastomers for Visible-Infrared-Radar Multispectral Camouflage. *Angewandte Chemie International Edition* **64**, e202422636 (2025).
68. Y. Wang, L. Zeng, J. Lin, Y. Qian, Z. Luo, F. Huang, L. Chen, Transparent programmable actuators based on MXene polymer composites: An approach to amphibious camouflage robots. *Chemical Engineering Journal*, 164157 (2025).
69. S. Ma, P. Xue, C. Valenzuela, Y. Liu, Y. Chen, Y. Feng, R. Bi, X. Yang, Y. Yang, C. Sun, 4D-Printed Adaptive and Programmable Shape-Morphing Batteries. *Advanced Materials*, 2505018 (2025).
70. J. Prat-Camps, G. Christopoulos, J. Hardwick, S. Subramanian, A manually reconfigurable reflective spatial sound modulator for ultrasonic waves in air. *Advanced Materials Technologies* **5**, 2000041 (2020).
71. J. Shintake, V. Cacucciolo, D. Floreano, H. Shea, Soft robotic grippers. *Advanced materials* **30**, 1707035 (2018).
72. H. Takahashi. (NHK, 2024).
73. E. W. Hawkes, C. Majidi, M. T. Tolley, Hard questions for soft robotics. *Science robotics* **6**, eabg6049 (2021).

74. N. Obayashi, D. Howard, K. L. Walker, J. Jørgensen, M. Gepner, D. Sameoto, A. Stokes, F. Iida, J. Hughes, A democratized bimodal model of research for soft robotics: Integrating slow and fast science. *Science Robotics* **10**, eadr2708 (2025).
75. N. Nagai, T. Matsunobe, T. Imai, Infrared analysis of depth profiles in UV-photochemical degradation of polymers. *Polymer degradation and stability* **88**, 224-233 (2005).
76. X. Pan, R. C. Verpaalen, H. Zhang, M. G. Debije, T. A. Engels, C. W. Bastiaansen, A. P. Schenning, NIR–vis–UV Light-Responsive High Stress-Generating Polymer Actuators with a Reduced Creep Rate. *Macromolecular Rapid Communications* **42**, 2100157 (2021).
77. G. Romero-Sabat, L. A. Granda, S. Medel, Synthesis of UV-curable polyurethane-acrylate hybrids with tuneable hardness and viscoelastic properties on-demand. *Materials Advances* **3**, 5118--5130 (2022).
78. K. Iiyoshi, S. Khazaaleh, A. S. Dalaq, M. F. Daqaq, G. Korres, M. Eid, Origami-based Haptic Syringe for Local Anesthesia Simulator. *IEEE Transactions on Haptics*, 1-6 (2024).
79. S. Chemicals, SYNTHON Chemicals Shop | (4-But-3-enyloxy)benzoic acid 4-methoxyphenyl ester | Liquid crystals reactive mesogens crown ethers calixarenes. *Synthon*, https://shop.synthon-chemicals.com/en/LIQUID-CRYSTALS/PHENYLBENZOATES/4-But-3-enyloxy-benzoic-acid-4-methoxyphenyl-ester.html.
80. Merck-KGaA, Polyacrylamide nonionic water-soluble polymer 9003-05-8. *Sigma-Aldrich*, https://www.sigmaaldrich.com/AE/en/product/sigma/92560.
81. Merck-KGaA, Poly(N-isopropylacrylamide) Mn 40000 25189-55-3. *Sigma-Aldrich*, https://www.sigmaaldrich.com/AE/en/product/aldrich/535311.

## Acknowledgments

We thank the Smart Materials Lab for providing MPTA fabrication and characterization resources. We thank Loubna Kahramane for collecting the actuator comparison data and Zuzanna Bak for the material characterization literature review and data analysis. We thank the NYUAD Core Technology Platforms, funded by Tamkeen, for providing cleanroom and SEM facilities for actuator fabrication and characterization. Google Gemini was used through New York University's GenAI service to polish the article's scientific writing.

## Funding:
New York University Abu Dhabi Center for AI and Robotics under the Research Institute Award (CG010)
New York University Abu Dhabi Global PhD Fellowship

## Author contributions

Conceptualization: KI, GK, SS, ME
Data curation: KI, GK, GR, ON, SS
Methodology: KI, GK, SS
Investigation: KI, GR, ON, SS
Visualization: KI, GK, GR
Software: KI, GK
Supervision: GK, SS, PN, ME
Writing—original draft: KI, GK, GR
Writing—review & editing: KI, GK, SS, PN, ME

## Competing interests

The authors declare that they have no competing interests.

**Correspondence** and requests for materials should be addressed to Ken Iiyoshi and Mohamad Eid.

## Supplementary materials:

Supplementary Text
Supplementary Figs. 1 to 6
Supplementary Movie 1 to 4
Source Data 1